\documentclass[11pt]{article}
\usepackage{amsmath,amssymb,amsfonts}
\usepackage{algorithmic}
\usepackage{graphicx}
\usepackage{textcomp}
\usepackage{xcolor}
\usepackage{enumitem}
\usepackage{bigints}
\usepackage{wasysym}
\usepackage[round,authoryear]{natbib}
\usepackage{pgfplots}
\usepackage{authblk}
    
\def\BibTeX{{\rm B\kern-.05em{\sc i\kern-.025em b}\kern-.08em
    T\kern-.1667em\lower.7ex\hbox{E}\kern-.125emX}}
        
\usepackage{fancyhdr}
\usepackage{tikz}
\usepackage{tikz}
\usepackage{pgfplots}
\usepackage{tikz-3dplot}

\usetikzlibrary{arrows.meta, shapes, calc, arrows, intersections, backgrounds, fit, positioning, bayesnet, snakes, decorations.markings}
\usetikzlibrary{shapes.geometric}
\usepgfplotslibrary{fillbetween}

\newcommand{\commentout}[1]{}
\newcommand{\pagesx}[1]{} 

\newcommand{\freee}[1]{}  
\newcommand{\newresultsout}[1]{}
\newcommand{\alloldresultsout}[1]{}
\newcommand{\allolderresultsout}[1]{}
\renewcommand{\cite}{\citep}
\usepackage{hyperref}

\begin{document}

\title{Classifying Problem and Solution Framing in Congressional Social Media}

\author[1]{Misha Melnyk} 
\author[1]{Mitchell Dolny}
\author[1]{Joshua D. Elkind}
\author[1]{A. Michael Tjhin}
\author[1]{Saisha Chebium}
\author[1]{Blake VanBerlo}
\author[2]{Annelise Russell}
\author[3]{Michelle M. Buehlmann}
\author[1]{Jesse~Hoey}
\affil[1]{David R. Cheriton School of Computer Science, University of Waterloo}
\affil[2]{Martin School of Public Policy \& Admin, University of Kentucky}
\affil[3]{Nistler College of Business and Public Administration,
University of North Dakota}



\maketitle

\begin{abstract}
  Policy setting in the USA according to the ``Garbage Can'' model differentiates between ``problem'' and ``solution'' focused processes. In this paper, we study a large dataset of US Senator postings on Twitter (1.68m tweets in total). Our objective is to develop an automated method to label Senatorial posts as either in the problem or solution streams. Two academic policy experts labeled a subset of 3967 tweets as either problem, solution, or other (anything not problem or solution). We split off a subset of 500 tweets into a test set, with the remaining 3467 used for training. During development, this training set was further split by 60/20/20 proportions for fitting, validation, and development test sets. We investigated supervised learning methods for building problem/solution classifiers directly on the training set, evaluating their performance in terms of F1 score on the validation set, allowing us to rapidly iterate through models and hyperparameters, achieving an average weighted F1 score of above 0.8 on cross validation across the three categories using a BERTweet Base model.
\end{abstract}

\section{Introduction}
Understanding complex political systems is a significant challenge that has occupied political scientists since time immemorial. Policy adoption in the United States is highly complex, involves many actors and many different processes, and extends over vastly different time ranges from days to decades. Two important elements of such systems are agendas and alternatives. Much early political science work lumps these together into a single ``agenda setting'' category, but~\cite{Kingdon2003} upended that concept. Through detailed ethnographic study, Kingdon's research unveiled that agendas and alternatives may actually follow quite different trajectories in policy setting. These trajectories differ in which participants are involved, which political processes or systems are present, and also a fair number of pragmatic factors and serendipity. It is this differentiation between agendas and alternatives that we approach in this work.

Kingdon opens his book with a number of examples, the first of which involves the creation of Health Maintenance Organizations (HMOs) in the early 1970s. The problem of medicare and medicaid is a perennial issue on the political agenda in the USA, involving both significant budgetary factors and partisan beliefs. A number of solutions were ``floating around'' in the 1970s, but none of them seemed to really fit the multifaceted healthcare problem, and the issue festered in the halls of congress. However, one actor (head of a policy or lobbying group) had come across this (currently well used) idea of private, competitive, health insurers. This particular actor happened to sit next to a very senior member of the Nixon administration on a plane, and, listening to him describe the federal medicare problems led him to propose the Health Maintenance Organization (HMO)  idea which moved forward to policy within a few weeks from that chance encounter. 

The HMO story is a classic example of a ``garbage can'' process~\cite{Cohen1972}, in which problems and solutions float around in a ``policy soup'' with political processes and actors. Sometimes, problems and solutions come together in some form of alignment, normally driven by some political entrepreneurs or by exogenous events. When this happens, a ``window of opportunity'' opens, through which a particular coupling of problem to solution floats through and into the public policy space for enactment (which is a third component).

Data describing these political processes is somewhat difficult to assemble, as it exists across so many different platforms, and so much of it is ``hidden'' (i.e. the conversations in the halls of congress). However, we do have a public-facing source of political behaviors in the form of social media posts by politicians. In this study, we examine a set of 1.68 million ``Tweets'' sent by US senators from January 2008 to February 2023. Our goal is to label these tweets as one of three categories: problems, solutions, or other. While the last category does not correspond precisely to the third of Kingdon's streams (political), we are primarily interested in distinguishing between tweets about problems and tweets about solutions. Once this dataset is labeled, it can serve as a valuable resource for investigation of these complex processes using a Kingdonian lens. In this paper, we focus on the technical aspects of confidently labeling this large dataset. 

\section{Datasets and Labeling}
This study analyzes public communications produced by members of the United States Senate on social media. Using the Twitter API, tweets were collected from official accounts associated with sitting U.S. Senators, producing a corpus of approximately 1.6 million tweets posted between 2008 and 2023. The dataset captures public-facing political communication through which elected officials identify policy concerns, promote legislative initiatives, and communicate institutional activity to constituents.
Each observation includes a unique tweet identifier, tweet text, author identifier, and posting date. During corpus construction, non-English tweets were excluded to ensure linguistic consistency for downstream modeling. Retweets and tweets containing embedded links were retained in order to preserve communicative context, reflecting how senators publicly engage with issues and audiences on the platform. The resulting corpus provides a large-scale record of elite political discourse suitable for examining how policymakers publicly frame policy problems and solutions over time.

\subsection{Human Labelling}
This study examines problem orientation on Twitter among U.S. senators. Given the Senate’s historical emphasis on individual autonomy, evidence of individualized problem-solving dynamics is especially revealing — and a key reason for focusing this analysis on the Senate. In addition, compared with the House, fewer individuals in the Senate make the analysis and initial hand coding of this dataset more feasible for classification. Using the Twitter API, we collected all tweets posted by official accounts linked to senators between January 2008 and February 2023. Each tweet contains a unique identifier (tweet\_id), the tweet’s text, the author’s identifier, and the date it was posted.

We first annotated a training set of 4203 tweets.  Two authors (MB and AR) were responsible for this, both experts in political science and the Kingdon model. They independently labeled each tweet  as ``problem,'' ``solution,'' or ``other'' and their ratings achieved an internal consistency of over 0.83 (Cohen's Kappa), and an F1-macro over $0.89$.  To understand senators’ problem-solving rhetoric, an initial random sample of tweets from 2017,  2019, and 2020, then a  stratified random sample of tweets from years 2012 to 2016, 2018 and 2021 were added for a total of 4203 tweets, which were expertly coded for the presence of problem or solution mentions. Tweets were first coded in a categorial fashion based upon whether the message includes any 1) problem-oriented or 2) solution-oriented language or 3) lacks either component. Policy problem rhetoric captures how senators use Twitter to highlight issues, challenges, or shortcomings in governance. Messages were coded as policy problem mentions if they explicitly identified a social, economic, or political issue in need of attention or reform. These tweets range from broad problem statements ({\em ``families are struggling to afford child care''}) to more specific critiques of existing policy ({\em ``SNAP funding needs to be increased''}). Such messages often serve as a foundation for senators' policy agendas, as the examples below illustrate.

{\em When we have nearly 60,000 people a month trying to cross the border illegally, climbing over old fences, our country is being assaulted. It shouldn't be hard for Democrats to agree with us on securing the border.''} - Bill Cassidy 2019/01/03

{\em A Nevada homeowner paid her mortgage on time for 15 years; then she lost her job and was injured in a car accident. @WellsFargo broke the law by refusing to work with her to save her home. This is unacceptable and I'm going to hold them accountable.} - Catherine Cortez Masto 2019/01/03

{\em .@SpeakerPelosi is right: climate change is an existential threat. With the new Congress seated, we can get to work on bold legislation like the \#GreenNewDeal. I look forward to working w/ the Speaker \& all of my colleagues who recognize the urgent need for transformative action.} – Richard Blumenthal 2019/01/03 

Policy solution rhetoric includes messages that emphasize proposed actions, reforms, or achievements aimed at addressing specific policy issues. Tweets were coded as policy solution mentions when they highlighted concrete steps, such as introducing legislation, implementing programs, or celebrating successful policy outcomes. These messages often underscore a senator’s role in advancing solutions or align their work with broader policy goals (“I introduced a bill to lower prescription drug costs for seniors”). In many cases, such tweets serve to demonstrate responsiveness and action, as the examples below illustrate (links removed):

{\em Nevada is leading the charge as the Innovation State with the responsible use of drones. My Drone Safety Enhancement Act, included in the 2019 FAA reauthorization bill, will increase safety and encourage the continued development of drone technology.} - Catherine Cortez Masto 2019/01/03

{\em His successor should start righting these wrongs beginning with approving the amendment to the Mashantucket Pequot casino compact, as required under federal law.} - Richard Blumenthal 2019/01/03

{\em Today I reintroduced the EL CHAPO Act. The bill would reserve any amounts forfeited to the US gov as a result of criminal prosecution of El Chapo \& other drug lords for border security assets and the completion of the wall along the US-Mexico border.} - Ted Cruz 2019/01/03

The non-problem/solution tweets — typical non-policy tweets altogether - are those messages that have no identifiable mentions of either a problem/solution. These messages are most commonly birthday messages, greetings, and mentions of constituent meetings.
To sum up, tweets were coded in a categorical fashion:
\begin{itemize}
  \item {\bf Problem (Category 1)} — messages that identify a social, economic, institutional, or political condition as problematic and in need of attention or reform (e.g., {\em ``Families are struggling to afford child care.''} {\em ``SNAP funding needs to be increased.''}).
\item {\bf Solution (Category 2)} — messages that emphasize proposed actions, reforms, concrete legislative steps, or achievements aimed at addressing a given policy problem (e.g., {\em ``I introduced a bill to lower prescription drug costs for seniors.''} {\em  ``We secured funding for wildfire mitigation.''}).
\item {\bf Other (Category 3)} — messages with no identifiable problem or solution content. These include ceremonial posts (e.g., birthdays, greetings), constituent interactions, event announcements, travel, and generic political messaging (e.g. {\em ``Allow me to welcome all of the new members who were just sworn in for the first time as senators and my friends who were re-elected to this body.''})
  \end{itemize}

Coding guidelines and additional examples are included in Appendix A. For these 4,203 tweets the two raters achieved 89.44\% agreement on TRAIN (89.8\% on TEST), with a Cohen’s Kappa of 0.838 (0.839 on TEST, weighted Kappa 0.843).  F1-scores computed for MB against AR were F1-macro 0.8943  and F1-weighted 0.8959. Based on~\cite{LandisKoch1977}, this is considered ``almost perfect agreement.'' 


\subsection{Data Processing}
When preparing the 4203 tweets labelled by human coders, the format provided resulted in the tweet IDs being truncated and some unicode being corrupted. To restore the ID labels we fuzzy matched the labeled tweets against the main tweets process to the larger dataset and successfully restored 3,967 tweet IDs with a score of at least 0.80. The majority of the tweets (3207) had a perfect match score of 1 and the rest with a score above 0.80 (760) were verified manually for correct matching. Of the 4,203 original labelled tweets, 236 labelled tweets with a lower score (0.80) were discarded as unreliable through this process. In addition, 1 tweet was coded as a ``4'' so it was removed. This incorrect tweet was from 2019.  The data set from this point forward is 3,966 labeled tweets  and 1,678,059 unlabelled tweets for a total of 1,682,025 tweets. We denote these sets as LABEL and UNLABEL respectively.

\subsection{Data Partitioning}

The labelled tweets were divided into two groups.  The first group was the TEST set of 499 labelled tweets. We intended this group to be 500 but the 1 mislabelled tweet discussed above was in this group, therefore it became 499. These were stratified so that there are 50 tweets from each year from 2012 to 2021, except 2019 which has 49. This set is not touched during our algorithmic development and will be the basis of our evaluation. The second group of tweets contained 3,467 labelled tweets and formed the TRAIN set.

Intercoder reliability remained high after these adjustments. On the main annotated set (LABEL, N = 3,966), raw agreement was 89.64\% with $\kappa = .840$ (weighted $\kappa = .846$). On the held-out test set (TEST, N = 499), raw agreement was 89.58\% with $\kappa = .840$ (weighted $\kappa = .832$).  In the TRAIN set, there are 359 tweets where AR and MB did not agree on all their labels. A subset of TRAIN was taken, TRAIN\_AGREE, containing 3108 tweets where both coders agreed. All label proportions are described in Table~\ref{tab:labels}.

\begin{table}[htb]
  \begin{center}
    \begin{tabular}{|l|c|c|c|c|c|c|}
      \hline
      Category & AR TRAIN & \% & MB TRAIN & \% & TRAIN\_AGREE & \%\\ \hline
      1 & 1337 & 39 & 1268 & 37 & 1156 & 37 \\
      2 & 743 & 21 & 747 & 21 & 638 & 21 \\
      3 & 1387 & 40 & 1452 & 42 & 1314 & 42 \\ \hline
    \end{tabular}
    \end{center}
    \caption{\label{tab:labels} Label counts and proportions for coded labels and agreed upon labels}
    \end{table}

  During development, TRAIN is split into DEV\_FIT, DEV\_VALIDATE, and DEV\_TEST sets in a 60/20/20 split. The training is performed on the labels by AR; this was chosen arbitrarily. This split is stratified on the label column so each set will have a similar proportion of each class. DEV\_FIT is used for fitting the model on the labeled training set. The idea is to use the DEV\_FIT data to learn a predictive model which we hope will generalize to the remainder of the dataset. DEV\_VALIDATE is used to check the model against data that it was not fitted against to perform early stopping of the training algorithm. Early stopping is a well known method to regularize and prevent overfitting. DEV\_TEST was used to evaluate the model and make decisions when iterating development.

\begin{figure}[htb]
  \begin{center}
    \includegraphics[width=\textwidth]{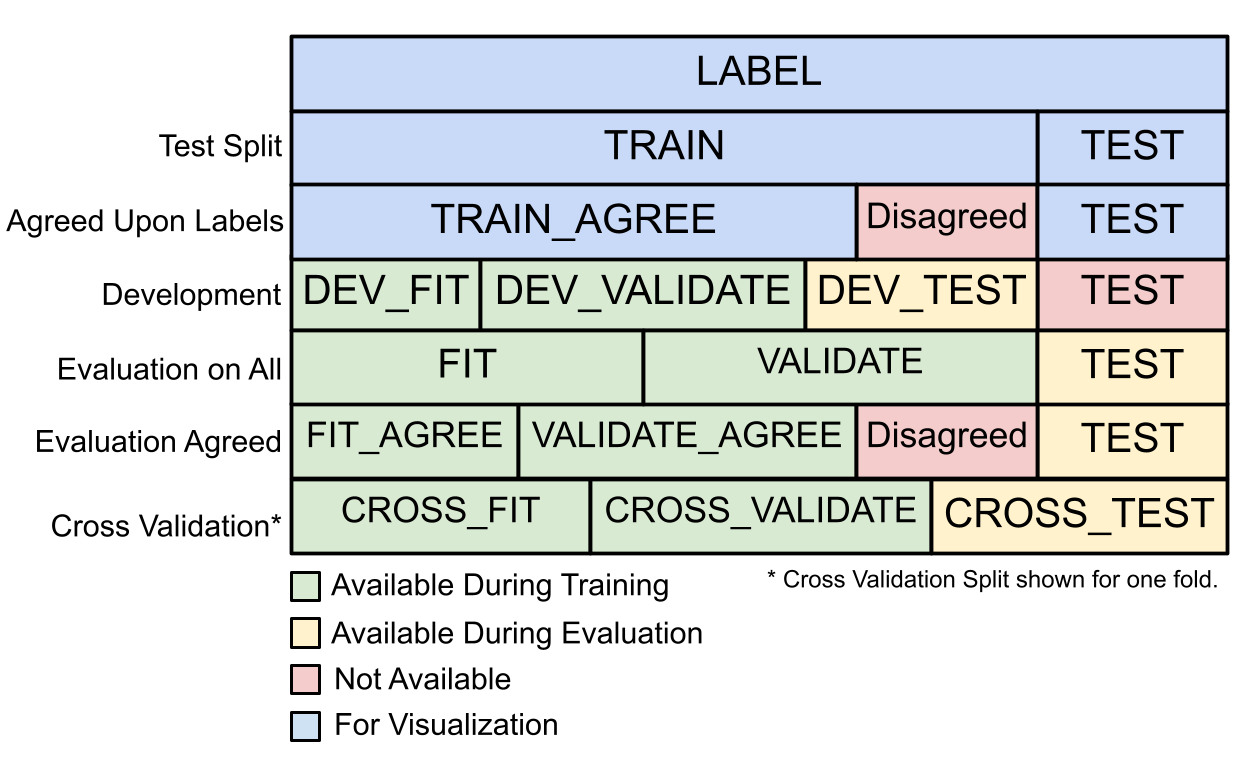}
    \end{center}
  \caption{\label{fig:fig1}Labeled data set partition used for training and evaluation. Size of subsets in graph are not indicative of set size.}
  \end{figure}

During evaluation, TRAIN was split into FIT and VALIDATE in an 80/20 split. Similarly, TRAIN\_AGREE was split into FIT\_AGREE and VALIDATE\_AGREE in an 80/20 split.  We also created data splits for k-fold cross validation with k = 7 (chosen arbitrarily). This uses all 3967 tweets in LABEL. For each fold, 1/7 of LABEL was split into CROSS\_TEST and the remaining 6/7 into CROSS\_TRAIN. A 80/20 split of CROSS\_TRAIN is used to create CROSS\_FIT and CROSS\_VALIDATE for the fitting and validation sets. CROSS\_TEST has no tweets in common across the 7 folds, i.e. the 7 folds tested against are disjoint. Figure~\ref{fig:fig1} is a schematic to visually display the various data splits used.

\section{Methods}
To construct a reliable large-scale classification pipeline, we conducted a two-stage experimental evaluation. In the first stage, we compared multiple classification approaches to determine which model architecture most effectively reproduced expert tweet annotations. In the second stage, we evaluated alternative training strategies for the selected classifier to identify the procedure that yielded the strongest predictive performance. The following sections describe the candidate classifiers, training procedures, and evaluation framework used in this process.

\subsection{Classifiers}
In the first stage of the experimental evaluation, we compared several classifier architectures to determine which modeling approach most effectively reproduced expert tweet annotations. We looked at three kinds of models: Logistic Regression~\cite{Bishop2006}, XGBoost trained decision trees~\cite{ChenGuestrin2016}, and BERT architecture neural networks~\cite{BERT2018}. These models either give a prediction directly or by outputting a confidence score. A confidence score assigns a value between 0 and 1 to each label, analogous to a probability. The sum of the confidence scores is 1.

Logistic regression takes real-valued points, applies a linear transformation, and applies a softmax operation to get confidence scores~\cite{Bishop2006}. A decision tree, such as with XGBoost~\cite{ChenGuestrin2016}, traverses a tree where each branch corresponds to some value present in the tweet. Both these methods use the year of the tweet and the author's numerical identifier. The BERT architecture is a transformer-based architecture introduced by researchers at Google~\cite{BERT2018}. Transformers are currently the state of the art models for natural language processing. They turn text into tokens using a tokenizer, either whole words or fragments of words, which are then transformed into high dimensional vectors called embeddings. The embedding process also accounts for where the token appears in the sentence, called positional embedding. These vectors are passed through several layers of neural networks to get final representations of the tokens.

Each layer calculates the importance of each token to the other tokens, called Attention. This gives a value from 0 to 1 of importance of each token in the sequence, with all importance values summing to 1. BERT models will have all tokens in each layer be influenced by all others. This results in transforming the original sequence based on this importance. The transformer starts all text with a special start token called Beginning of Sequence (BOS). Since all tokens influence each other, the final embedding of the BOS is treated as a representation of the whole text.

\begin{figure}[htbp]
  \begin{center}
    \includegraphics[width=\textwidth]{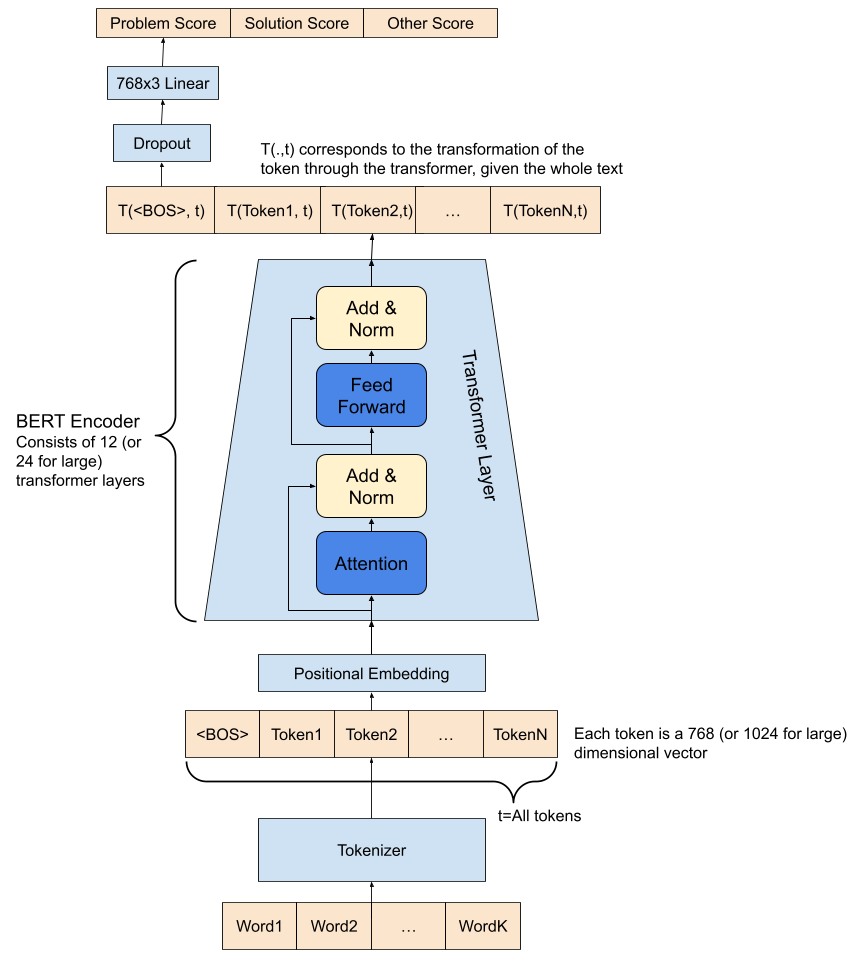}
    \end{center}
  \caption{\label{fig:fig2} BERT Model Details showing flow of data from initial text to output scores. Information flows upwards in this picture, starting from a sequence of words (the Tweet) at the bottom. The Tokenizer separates out the raw text into tokens which are vectors in an embedding space. These vectors are augmented with positional information and fed into the Bidirectional Transformer (shown summarized as an attention layer and a feed forward layer in blue). The outputs of this model are then fed through a dropout and a linear layer to produce the final classification into problem/solution/other.}
  \end{figure}

Next, a dropout layer is applied to this embedding. During training, it randomly zeroes out some values in the embedding and scales the rest, while not doing so during evaluation. This is meant to make each value in the embedding generalize more. For classification, the output of the dropout is fed to a final smaller neural network to produce confidence scores from the representation of the whole sentence. Figure~\ref{fig:fig2} is a schematic of the model showing the various layers.

Two variants of BERT are used: BERT base and BERT large. The difference between them is the number of layers and dimensionality of the embeddings, shown in Table~\ref{tab:weights}. As a result, BERT large has more parameters and can store more information within each embedding.

\begin{table}
  \begin{center}
    \begin{tabular}{|l|l|l|}
      \hline
    & BERT-base & BERT-large \\ \hline
    Total Weights & 110 million & 340 million \\
    Attention layers & 12 & 24 \\
    Embedding Dimension & 768 & 1024\\ \hline
  \end{tabular}
  \end{center}
  \caption{\label{tab:weights} BERT base vs large characterists comparison}
\end{table}

\subsubsection{Loss Function and Optimizer}
The loss function and optimizer are two choices that are made for training a model. The loss function is the objective you want to try to optimize for, trying to minimize the loss. The optimizer defines the update step for the model parameters. The training uses the Cross Entropy loss and AdamW optimizer which are the current standard for classification.

Cross entropy loss is designed specifically for classification. We require the following: Tweet $x$, Label $t$, and BERT model $M$. The tweet is run through the model to get a score for each category: $y = M(x)$. These scores are converted into a probability vector through the softmax function: $p = Softmax(y)$, where $p_i = \frac{e^{y_i}}{\sum_{k=1}^{3} e^{y_k}}$ represents the confidence of the model that the tweet is class $i$. Therefore $p_t$ will represent the confidence that the model predicts for the correct class  $t$. The label will be one of the values 0, 1, or 2 corresponding to the Problem, Solution, or Other classes. The label will be transformed into a one-hot vector $q$ where $q_t = 1$ and all other entries are 0. The cross entropy loss is then
$L(p,q)= -\sum_{i=1}^{3}q_i\log(p_i)$. With the one-hot vector, this becomes $-\log(p_t)$. The goal of training will be to minimize this value for all tweets. In order to minimize $-\log(p_t)$ we want to maximize $p_t$, therefore, changing the parameters to reduce this loss should improve the classification of the tweets.

The AdamW optimizer~\cite{LoshchilovHutter2017} performs Adaptive Momentum updates with decoupled weight decay. For each iteration of training, the AdamW optimizer calculates the gradients with respect to the loss of the training example. These are used for the exponential averages of variables mt and vt for the adaptive momentum portion. The parameters $\beta_1$ and $\beta_2$ are used for adaptive momentum and are by default $0.9$ and $0.999$. For gradient $g$, $m_t= \beta_1 m_{t-1} + (1-\beta_1)g$, and $v_t$ uses $\beta_2$ and $g^2$. They are normalized by a factor of $\frac{1}{(1-\beta_1^t)}$ (c.f. $\beta_2$) to get $m_t'$ and $v_t'$. AdamW also performs weight decay directly by the originating formula in the update step. The idea behind weight decay is to regularize the parameters by scaling their magnitude each iteration by a parameter $\lambda$. The difference between AdamW and Adam is that Adam does this through L2 Regularization, which is equivalent to the scaling for other methods, but not for Adaptive Momentum.

The equation for the update step becomes the following with learning rate $\alpha$ and $\epsilon=10^{-8}$ and $\eta_t = 1$:
\[\theta_t= \theta_{t-1} - \eta_t(\alpha\frac{m_t}{\sqrt{v_t+\epsilon}} + \lambda \theta_{t-1})\]
The  is there to prevent division by 0. The ratio $\frac{m_t}{\sqrt{v_t}}$ is meant to serve as a Signal-to-Noise ratio~\cite{Kingma2014adam}. $m_t$ serves as the direction for the parameters to update. $v_t$, being the exponential average of the gradients squared, is the uncentered variance~\cite{Kingma2014adam} and is used as a measure of how noisy the gradients are. The noisier the gradients, the less we trust the momentum, leading us to take a smaller step. In more stable regions where the noise is low, $\sqrt{v_t+\epsilon}$ will be smaller so we can take larger steps.

\subsubsection{BERT Training Process}
There are two kinds of training experiments that we did to fine-tune the BERT model for our classification task (see Section 2.3). The first is only using subsets of  TRAIN as examples for fitting and validation. The second is k-fold cross validation, using CROSS\_TRAIN and CROSS\_VALIDATE This involves creating k CROSS\_TEST sets and splitting the rest into CROSS\_TRAIN and CROSS\_VALIDATE. Training is performed k times, once for each fold, and k is shown in Table 3 as the cross\_val\_folds parameter, where we used k=7. 

The BERT training uses a standard supervised learning setup with training examples. Datapoints contain text from the tweet and the class label; publish date and author are not used, such that the determination is made purely on the tweet itself. The text is run through the transformer encoder, then the BOS token is used in a dropout and a single linear layer to get scores for each label. These scores are trained with a Cross Entropy loss using the coded label and the AdamW optimizer. The loss is then used to determine how to adjust the model weights through backpropagation in order to make the model predict the coded label with higher confidence for the given tweet. Fitting is also scaled by the learning\_rate parameter.. The BERT transformer starts off with pretrained weights given by vinai/bertweet-base (Nguyen et al., EMNLP 2020), specified as model\_name. The dropout chance is controlled by a parameter dropout\_p, which is set to 0.1.

Training consists of multiple training epochs. One epoch consists of using every datapoint once for fitting. Several tweets are looked at at once, called a batch. The number of tweets in each batch corresponds to the batch\_size parameter in Table 3. In many applications, the model weights are updated after every batch. Here though, we added the option to perform the fitting every few batches. This corresponds to the accumulate\_grad\_batches parameter in Table 3.

After each training epoch, the VALIDATION  is used to evaluate the performance of the model. The model parameters are not fit to the tweets in VALIDATION. The metrics evaluated on VALIDATION are:

\begin{itemize}
  \item Accuracy - Proportion of validation set tweets predicted correctly.
\item F1-Score - The average F1-Score metric on each label (F1-score is calculated on each of the three classes, then divided by the number of classes. i.e. Macro average)
\item Weighted F1-Score - The weighted average F1-Score, weighted by number elements in each class.
\item Confusion Matrix - A table of predicted labels vs. true labels on two axes.
  \end{itemize}

These metrics are tracked across epochs. After the validation step is complete, an early-stopping check is performed. The purpose of early stopping is to prevent overfitting. The early stopping performed here is checking the increase of the F1-Score. If it hasn’t gone up for several epochs on VALIDATION (defined in the stopping\_patience parameter), then we consider the model to have started to overfit. From here, we take the model that had the highest F1 metric on VALIDATION and evaluate it on the corresponding TEST set.

This process we refer to as a Trial. In summary, it includes the following:
\begin{itemize}
\item Randomly splitting the training set into fitting and validation sets.
\item Fitting the model for an epoch, passing over the whole fitting set once.
\item Evaluating the model on the validation set.
\item Repeating the above two steps for max\_epochs or until overfitting heuristics are satisfied.
\item Evaluating the model on the test set.
\end{itemize}
Each trial will have an element of pseudo-randomness in three parts:
\begin{itemize}
\item The datapoints are shuffled in each trial
\item The initial parameters for the final linear layer are randomly initialized
\item The dropout during training is based on random values.
\end{itemize}

This pseudo-randomness is affected by an initial seed. We set the seed during each trial to global\_seed in Table~\ref{tab:hyper} plus the trial number it is. This will make each trial different while also making the whole experiment reproducible.

During development, we explored two ways of calling the BERT model. The first is SentenceTransformer\footnote{\url{https://www.sbert.net}~\cite{reimers2019}}. This takes in the text directly, performs tokenization, and runs it through the BERT model. The final linear layer is done manually on top of this embedding. The second is with AutoModelForClassification by HuggingFace,\footnote{\url{https://huggingface.co/docs/transformers/model\_doc/auto}} which takes tokens as input and outputs the confidence scores. We initially worked with the SentenceTransformer and performed a hyperparameter search with modified learning rate and dropout percentage to see which combination works best for our metrics. These did not reach to the same level that the AutoModelForClassification did, so we settled for the latter. One reason we discovered that might have caused this is that SentenceTransformers does not allow updating the BERT parameters as it is run in evaluation mode, which does not calculate the gradients needed for weight updates.

In total, 20 trials are run of the main evaluation training. 140 runs are done of cross validation, 7 folds in each of 20 trials.
The hyperparameters and configuration used is shown in Table~\ref{tab:hyper}. All other parameters were default, either defined by PyTorch or PyTorch Lightning. The full dataset and codebase is available at \url{cs.uwaterloo.ca/~jhoey/research/polytweets/}.

\begin{table}
  \begin{center}
    \begin{tabular}{|l|l|l|}
      \hline
      model\_name & vinai/bertweet-base1U  & The base model used \\
dropout\_p & 0.1 & Dropout probability in training \\
trials & 20 & Number of different starting seeds\\
cross\_val\_folds & 7 & How many folds for k-fold cross validation \\
learning\_rate & 0.00003 & Learning rate for optimizer \\
max\_epochs & 25& Maximum number of epochs to do before stopping \\
accumulate\_grad\_batches& 1& How many batches to do before updating gradients\\
stopping\_patience & 3 & Number of epochs before triggering early stopping \\
batch\_size & 64 & Number of tweets in a mini-batch \\
global\_seed & 2025 & Seed value for starting state \\ \hline
  \end{tabular}
  \end{center}
  \caption{\label{tab:hyper}. Parameters used while running the training trials.}
  \end{table}

\subsection{Training Strategy Evaluation}
In the second stage of the experimental evaluation, we evaluated alternative training strategies for scaling classification beyond the labeled dataset. While the classifier architecture remained fixed, these approaches differed in how labeled and unlabeled data were incorporated into the learning process. We compared pseudo-label–based self-training and large language model (LLM) prompting approaches to determine which training procedure produced the strongest predictive performance under a common evaluation framework.

Three unsupervised methods were attempted. Each of these has a mechanism for using data without labels. The first method trains a supervised model on the labeled data (as in the last section) and then uses it to assign labels with confidences to every unlabeled example~\cite{Frisli2025}. Those labels with the highest confidence are then used to fine-tune the model. The second two methods use large language model prompting, and require no training data at all.

Unsupervised learning has the advantage that fewer (or no) labels are needed. We explored two types of unsupervised learning: LLM-based and pseudo-label based. LLM-based methods use a pre-trained large language model and retrieve a classification for a tweet by prompting the LLM. We found that modern "AI" models were capable of classifying text according to Kingdon's model without having the model explained. We investigated using an LLM directly to classify each tweet as problem/solution/political (``direct method''), and also using it to give confidences in each of the three categories, which are then optimized over using the VALIDATION set (``confidence method''). 
We explore two LLMs: GPT-4.1-nano and GPT-4.0 via the Martian API. All results are shown for GPT-4.0, as it produced clearer, more consistent outputs.

We call these LLM based methods ``unsupervised'' except one must be aware that the LLM training data included (1) many documents and descriptions of the Kingdon model; and (2) the exact tweet being analyzed (very likely). Therefore, while it may not have seen an explicit "problem" label assigned to Senator A's tweet {\em ``Climate change is destroying our cities,''} it may well have seen both the tweet, and some text analyzing the tweet such as the following headline on Fox News: {\em ``Senator A said that climate change is a problem for our cities with his tweet...''} Thus the LLM may well have seen, in its dataset, an explicit label for this tweet. Even without such an explicit label, the LLM is excellent at learning wider dependencies, and so might already have learned that {\em ``climate change''} is a problem rather than a solution.

\subsubsection{Active Learning}
We use an iterative pseudo-labeling (self-training) approach~\cite{Frisli2025} based on a supervised BERTweet-base model with class weighting, trained on the 3,467 labeled tweets in TRAIN. Then, the remaining tweets in UNLABEL are given a label and a confidence between 0.0 and 1.0. Any tweet in UNLABEL with a confidence above threshold above a constant value c is then added to LABEL with the predicted category (in problem/solution/political) as a pseudo-label. We start training with the maximum possible confidence threshold (c=1.0), reducing it by 0.05 each iteration. If no predictions meet the current threshold, it is immediately decreased by a further 0.05. The minimum level is 0.7, a floor which prevents the inclusion of low-quality pseudo-labels. In our experiments, 0.7 provided an optimal balance between quality and quantity. For more sensitive domains, higher floors (e.g., 0.8 or 0.85) may be appropriate. 

At each iteration, the model is trained on the current labeled dataset. It then predicts labels and confidence scores for all tweets in UNLABEL. Predictions whose confidence exceeds a threshold c are added to the labeled dataset as pseudo-labels, subject to class balancing constraints. The model is retrained on this expanded dataset, and the threshold c is gradually reduced until a minimum floor is reached. All final results are for TEST.

An important aspect of the pseudo-labeling process is managing class imbalance, which can be exacerbated through iterative training. Majority classes (e.g., Class 1: Problem-Oriented) typically received higher confidence scores. For example, in Iteration 2 (after one round of pseudo-label addition), Class 1 had mean confidence of 0.94, while Class 3 had mean confidence of 0.87, which creates a risk of amplifying existing imbalances through pseudo-labeling. Thus, each iteration tracks the distribution of classes in selected pseudo-labels. For example from early iteration: \{Class 1: 58\%, Class 2: 32\%, Class 3: 10\%\}, while from a later iteration \{Class 1: 45\%, Class 2: 38\%, Class 3: 17\%\}. The gradual improvement in balance occurs as the model becomes more confident in minority classes. We therefore use class weighting in the loss function to counteract imbalance in both original and pseudo-labeled data. Further, higher thresholds for majority classes can be implemented (e.g., 0.85 for Class 1, 0.75 for Class 3).  In extreme cases, class-specific quotas can limit the number of pseudo-labels added per class.

Overall, we found the threshold parameters c significantly impacted how the training dataset grows. With a conservative approach (higher Floor, e.g., 0.85) there is slower growth in training data, but higher quality pseudo-labels. A balanced approach (moderate Floor, e.g., 0.7) leads to moderate growth in training data and good quality pseudo-labels with acceptable error rate. An aggressive approach (lower Floor, e.g., 0.6), there is rapid growth in training data, but higher risk of error propagation. 

\subsubsection{LLM based approach: Direct}
Our first method uses LLM prompting for direct tweet classification. We evaluate two prompt variants: one requiring a class label with a brief explanation, and one requiring only the class label. The following prompt was used for the Direct classification method with explanation included:

\begin{footnotesize}
\begin{verbatim}
Based on Kingdon's theory, please classify this tweet into one of these categories:
1. Problem Oriented - The tweet describes or mentions a problem, issue, or challenge
2. Solution Oriented - The tweet describes or mentions a solution, fix, or resolution
3. Political - The tweet is political in nature but doesn't clearly focus on problems or solutions
Respond with ONLY the number (1, 2, or 3) corresponding to the category, followed by a brief explanation.
Format your response as: NUMBER [explanation]
For example: 1 [This tweet focuses on describing a problem with healthcare costs]
Tweet: [tweet text]
\end{verbatim}
\end{footnotesize}

The following prompt was used for the Direct classification method without explanation included:

\begin{footnotesize}
\begin{verbatim}
Based on Kingdon's theory, please classify this tweet into one of these categories:
1. Problem Oriented - The tweet describes or mentions a problem, issue, or challenge
2. Solution Oriented - The tweet describes or mentions a solution, fix, or resolution 
3. Political - The tweet is political in nature but doesn't clearly focus on problems or solutions
Respond with ONLY the number (1, 2, or 3) corresponding to the category.
Tweet: [tweet text]
\end{verbatim}
\end{footnotesize}

The no-explanation variant achieved higher accuracy and macro F1, likely due to reduced output variability and lower generation noise. As a result, we report results for both variants but treat the no-explanation prompt as the primary direct-classification baseline.

\subsubsection{LLM based approach: Confidence}

The confidence-based k-threshold approach represents a novel application of generative AI for tweet classification. Unlike traditional direct classification, this method leverages LLMs' ability to express uncertainty through confidence scores, providing a more nuanced classification framework.

LLM provides explicit confidence percentages for Classes 1 and 2
Prompt engineering ensures scores are expressed as percentages (0-100
Class 3 confidence implicitly calculated as remaining percentage (100 - conf1 - conf2)

We use a two-stage confidence-based classification process: (1) confidence generation and (2) threshold-based decision making. First, we generate a confidence score by prompting the LLM  (see below for prompts) to give confidence percentages for Classes 1 and 2. The prompt ensures scores are expressed as percentages (0-100\%) and Class 3 confidence implicitly calculated so all three sum to 100\%. 
If the confidence in Class 3 is greater than a parameter k, then we classify it as class 3, otherwise it is classed as the higher confidence of Class 1 and 2. 

The following prompt was used for the Confidence based classification method with explanation included:

\begin{footnotesize}
\begin{verbatim}
Based on Kingdon's theory, please provide confidence scores for this tweet.
Classify this tweet into these categories:
1. Problem Oriented - The tweet describes or mentions a problem, issue, or challenge
2. Solution Oriented - The tweet describes or mentions a solution, fix, or resolution 
For this tweet, please provide:
- Confidence score for Class 1 (Problem) as a percentage (0-100)
- Confidence score for Class 2 (Solution) as a percentage (0-100)
Note: The sum of both confidence scores should not exceed 100%, may leave room for other.
Format your response as: CONF1,CONF2 [explanation]
For example: 85,10 Political discussion with some problem elements
Tweet: [tweet text]
\end{verbatim}
\end{footnotesize}

The following prompt was used for the Confidence based classification method without explanation included:

\begin{footnotesize}
\begin{verbatim}
Based on Kingdon's theory, please provide confidence scores for this tweet.
Classify this tweet into these categories:
1. Problem Oriented - The tweet describes or mentions a problem, issue, or challenge
2. Solution Oriented - The tweet describes or mentions a solution, fix, or resolution 
For this tweet, please provide:
- Confidence score for Class 1 (Problem) as a percentage (0-100)
- Confidence score for Class 2 (Solution) as a percentage (0-100)
Note: The sum of both confidence scores should not exceed 100%, may leave room for other.
Format your response as: CONF1,CONF2
For example: 85,10
Tweet: [tweet text]
\end{verbatim}
\end{footnotesize}

Second, we use a threshold-based classification method in which parameter k represents the threshold for Class 3 classification. That is, we look at the confidences of Class 3 and 
If implicit Class 3 confidence > k: classify as Class 3
Otherwise: classify as the class with higher confidence between 1 and 2

We then optimized the k-threshold using a grid search from k=1.0 to 100.0 in increments of 1.0.  Each threshold evaluated against DEV\_VALIDATE. Performance metrics were recorded for each k value. For low k values (e.g., 5-15) there was aggressive Class 3 assignment, high recall but low precision. For medium k values (e.g., 20-40), there was balanced performance across classes. For high k values (e.g., 45+) there was conservative Class 3 assignment, high precision but low recall. We optimized the threshold looking for maximum accuracy (optimal k typically between 25-35) and maximum macro F1 score (optimal k typically between 15-25). 

\section{Results}
This section reports results corresponding to the two-stage experimental evaluation described in Section 3. We first present results comparing alternative classifier architectures used to predict tweet labels. We then report results evaluating training strategies applied to the selected classifier, including iterative pseudo-labeling and large language model–based approaches. Together, these analyses identify the modeling configuration used to label the full dataset.

\subsection{Classifier Evaluation Results}
To identify the classifier architecture that most effectively reproduces expert tweet annotations, we compared several supervised learning approaches, including logistic regression, XGBoost decision trees, and transformer-based BERT models. Performance was evaluated using macro F1 and weighted F1 scores computed on held-out evaluation data.

\begin{table}[htb]
  \begin{center}
    \begin{tabular}{|l|l|l|l|l|l|l|}
      \hline
      Classifier & Logistic Regression & XGBoost Treest & \multicolumn{2}{c|}{BERT Base} & \multicolumn{2}{c|}{BERT Large} \\ \cline{4-7}
      & [AR] & [AR] & [AR] & [MB] & [AR] & [MB] \\ \hline
      Avg F1  & 0.44 & 0.42 & 0.813 & 0.796 & 0.800 & 0.836 \\ \hline
      Avg Weighted F1 & 0.47 & 0.45 & 0.806 & 0.823 & 0.810 & 0.845 \\ \hline
  \end{tabular}
  \end{center}
  \caption{\label{tab:compres} Performance comparison of classifier architectures on held-out evaluation data (macro F1 and weighted F1 scores).}
  \end{table}

\begin{table}[htb]
  \begin{center}
\begin{small}
    \begin{tabular}{|l|l|l|l|l|l|l|}
      \hline
      Variant & \multicolumn{3}{c|}{Base} &  \multicolumn{3}{c|}{Large} \\ \cline{2-7}
                     & agreed & disagreed  & x-val & agreed & disagreed & x-val \\ \hline
      AR m-F1          & 0.796 (0.0084) & 0.800 (0.0089)& 0.799 (0.016) & 0.800 (0.011)& 0.796 (0.011) & 0.806 (0.02)\\
      AR w-F1  & 0.806 (0.0084) & 0.810 (0.0084)& 0.812 (0.015) & 0.810 (0.0099) & 0.807 (0.01) & 0.817 (0.02)\\
      MB F1          & 0.813 (0.0093) & 0.810 (0.011) & -     & 0.836 (0.014) & 0.810 (0.013) & - \\
      MB m-w-F1 & 0.823 (0.0091) & 0.819 (0.010)& -     & 0.845 (0.013) & 0.820 (0.011) &\\ \hline
      AR accuracy & 0.789 (0.0083) & 0.79 (0.0070) & 0.800 (0.017) & 0.790 (0.013) & 0.789 (0.017) & 0.804 (0.02)  \\ 
      MB accuracy & 0.805 (0.0094) & 0.805 (0.011) & - & 0.807 (0.022) & 0.774 (0.019) & - \\ \hline
  \end{tabular}
\end{small}
  \end{center}
  \caption{\label{tab:fullres} Performance of BERT model variants across training configurations and validation settings (macro F1 and weighted F1 scores denoted m-F1 and w-F1, respectively). Standard deviations across 20 random trials shown in parentheses.}
  \end{table}

\commentout{
================
Bert-base

Here are accuracy numbers for Bert-base vs AR and MB
AR Avg. Accuracy:
Agreed: 0.7934, std 0.0093
Disagreed: 0.7927, std 0.0081
MB Avg. Accuracy:
Agreed: 0.8108, std 0.0074
Disagreed: 0.8002, std 0.0089

here are the std. deviations on the F1 scores
AR F1-Std:
Agreed Macro: 0.0095
Agreed Weighted: 0.0087
Disagreed Macro: 0.0093
Disagreed Weighted: 0.0088
MB F1-Std:

Agreed Macro: 0.0064
Agreed Weighted: 0.0063
Disagreed Macro: 0.0090
Disagreed Weighted: 0.0086

================
Bert-large

AR Accuracy:
Agreed - Mean 0.8112 Std 0.0096
Disagreed - Mean 0.8079 Std 0.0093

MB Accuracy:
Agreed - Mean 0.8460 Std 0.0126
Disagreed - Mean 0.8211 Std 0.0105

AR F1 Std:

Agreed Macro - 0.0108
Agreed Weighted - 0.0099
Disagreed Macro - 0.0105
Disagreed Weighted 0.0101
MB F1 Std:

Agreed Macro - 0.0144
Agreed Weighted - 0.0132
Disagreed Macro - 0.0133
Disagreed Weighted - 0.0116
=========
}

As shown in Table~\ref{tab:compres}, transformer-based models substantially outperform classical machine-learning baselines, with logistic regression and XGBoost achieving considerably lower macro and weighted F1 scores than BERT-based classifiers. Across evaluation settings, BERT models consistently achieve weighted F1 scores exceeding 0.80, substantially outperforming classical baselines. Accuracies are also similarly around 0.80. Standard deviations across 20 random trials are mostly near  $0.01$ on accuracies and F1 scores.  Logistic Regression and XGBoost Trees use just the year and author and serve as baselines, while BERT uses just the tweet text. This suggests that the text does have a signal that differentiates it between these classes. Table~\ref{tab:fullres} further examines performance within transformer architectures, demonstrating that results remain stable across alternative training configurations and validation splits. Performance is comparable across annotation variants, with the BERT base (BERTweet) configuration achieving the strongest and most consistent overall performance. Notably, the comparison with the evaluation on MB’s label’s is higher when trained on only agreed-upon tweets. This may be because the disagreed-upon tweets would not cause the model to optimize for patterns that are contrary to MB’s labelling. This would also explain why the evaluation with AR’s labels took a smaller hit with the disagreed-upon tweets included. Interestingly though, the evaluation on MB’s labels was in both instances on average higher than AR’s in both scenarios. Based on these results, the BERTweet (BERT base) model was selected as the classifier used in subsequent training-strategy experiments. Some examples are shown in Table~\ref{tab:examples}.

\subsection{Training Strategy Evaluation Results}
Following selection of the BERTweet classifier, we evaluated alternative training strategies to determine how best to scale classification beyond the manually labeled dataset. Specifically, we compared iterative pseudo-labeling (self-training) with large language model (LLM)–based classification approaches. Performance was evaluated using accuracy, macro F1, and weighted F1 scores computed on held-out evaluation data. The goal of this stage was to identify the training procedure that produced the strongest predictive performance for large-scale dataset labeling.

\begin{table}
  \begin{center}
    \begin{tabular}{|l|l|l|l|}
      \hline
      Training Strategy & Accuracy & Macro F1& Weighted F1\\ \hline
Iterative Pseudo-Labeling& 81.6\% & 0.807& 0.818\\
LLM Direct Classification (no explanation) & 65.8\% & 0.481 & 0.648 \\
LLM Confidence-Based Classification (k = 5) & 60.9\%  & 0.374 & 0.666 \\ \hline
  \end{tabular}
  \end{center}
  \caption{\label{tab:unsupres} Overall scores for unsupervised or semi-supervised methods.}
  \end{table}

As shown in Table~\ref{tab:unsupres}, iterative pseudo-labeling substantially outperforms both LLM-based approaches across all evaluation metrics. The pseudo-labeling strategy achieves the highest accuracy and macro F1 score, indicating stronger agreement with expert annotations when extending classification beyond the labeled dataset. In contrast, LLM-based approaches show substantially lower macro F1 scores, reflecting reduced performance in distinguishing among the three tweet categories. Based on these results, iterative pseudo-labeling was selected as the training strategy used for large-scale classification of the full dataset.

The best-performing pseudo-labeling configuration occurred at Iteration 2, using a confidence threshold of 0.94. At this stage, the training set expanded to 3,659 samples (3,467 labeled and 192 pseudo-labeled). This configuration achieved 81.6\% accuracy and a macro F1 score of 0.807, representing the strongest overall performance across all evaluated methods.
Performance degraded slightly in later iterations, suggesting diminishing returns and increased risk of error propagation as lower-confidence pseudo-labels were incorporated.

For direct LLM-based classification, the prompt variant without explanations achieved the best performance. This configuration yielded an accuracy of 65.8\% and a macro F1 score of 0.481.
Including explanations consistently reduced performance, likely due to increased output variability and generation noise.

The best-performing confidence-based configuration used the no-explanation prompt with a threshold of k = 5. This setting maximized overall accuracy (60.9\%) but resulted in a lower macro F1 score (0.374), reflecting reduced precision for minority classes. The LLM used was GPT-4.0 via the Martian API.

\section{Downstream statistics}
Having established a classification model capable of reliably reproducing expert annotations, we next present descriptive statistics derived from the fully labeled dataset. These analyses are exploratory and intended to illustrate the kinds of substantive questions that the dataset enables rather than to test specific hypotheses.

\begin{figure}[htb]
  \begin{center}
    \includegraphics[width=\textwidth]{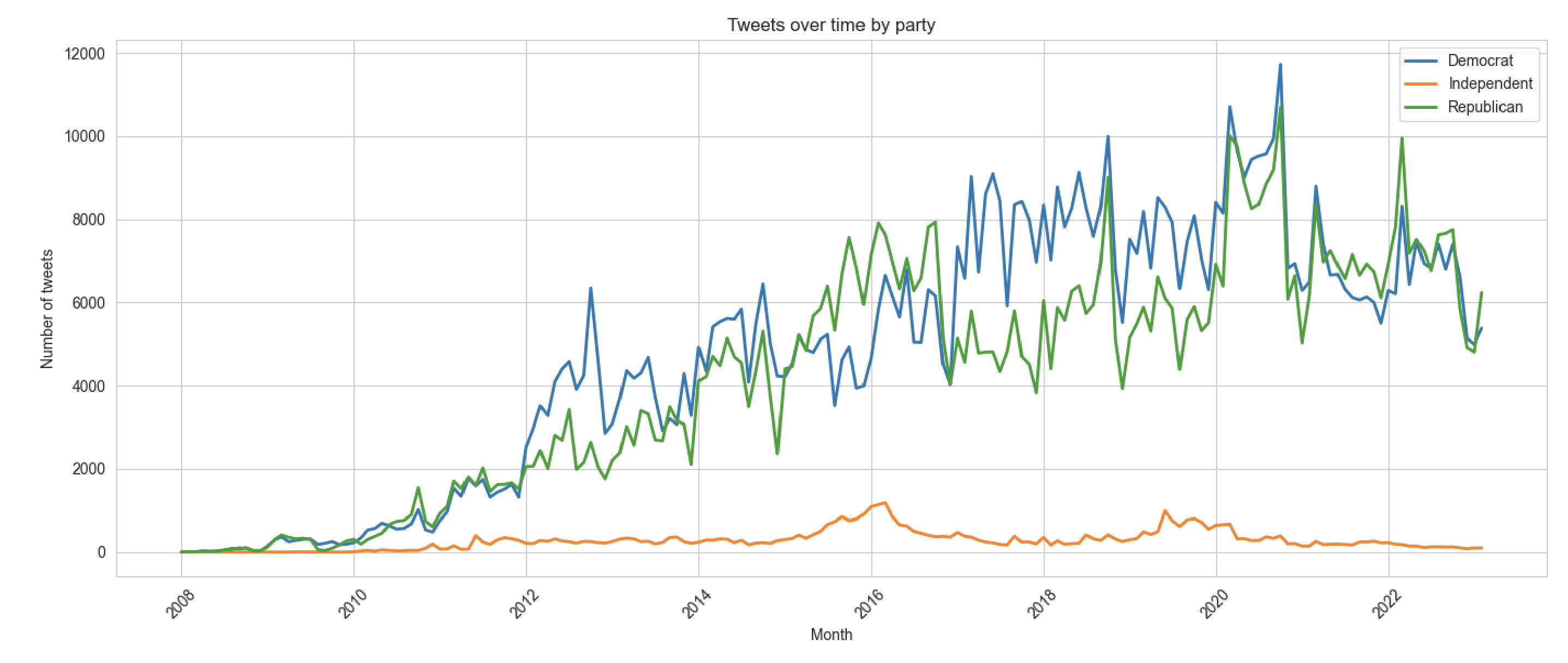}
    \end{center}
  \caption{\label{fig:fig4} Tweets per month by party}
  \end{figure}

Figure~\ref{fig:fig4} the number of tweets per month by party. We can see the steep rise starting between 2010-2012 when Twitter was growing in popularity.

\begin{figure}[htb]
  \begin{center}
    \includegraphics[width=\textwidth]{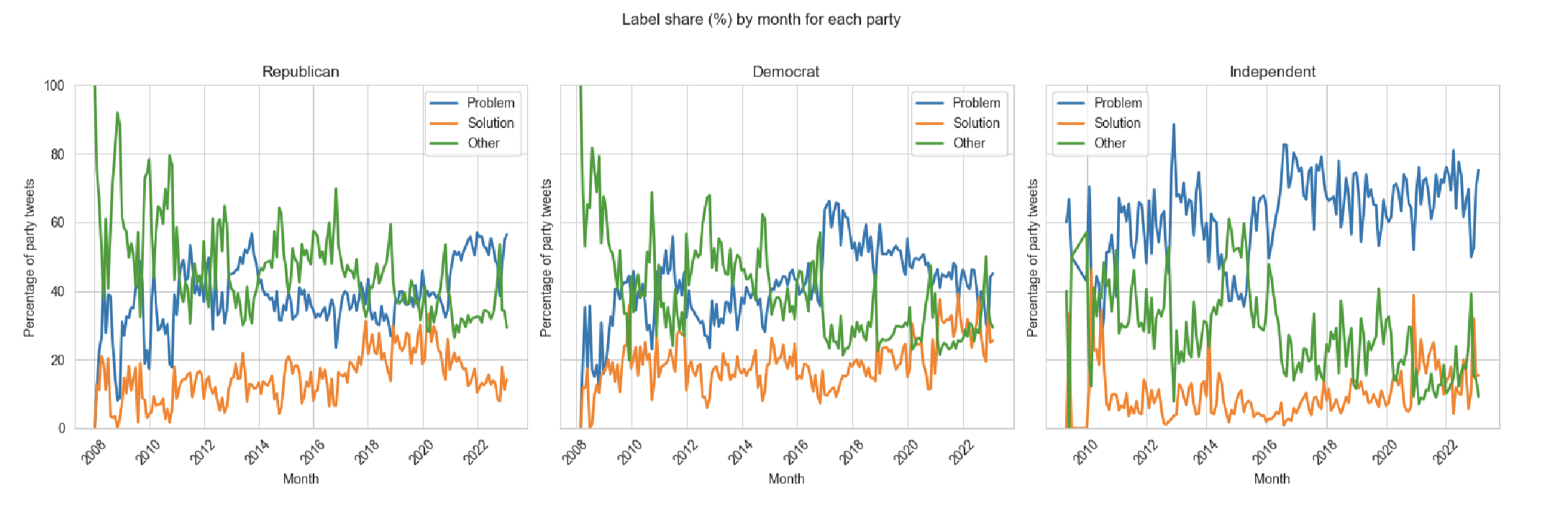}
    \end{center}
  \caption{\label{fig:fig5} Distribution of labels per month for each party.}
\end{figure}

Figure~\ref{fig:fig5} shows the distribution of labels per month for each party. One thing that stands out is the increase in democrat "problem" tweets in 2017, DJT first term, and the mirror image in republican tweets in 2021, Biden first term. Solution tweets also see a bit of a surge in 2017 perhaps.

\begin{figure}[htb]
  \begin{center}
    \includegraphics[width=\textwidth]{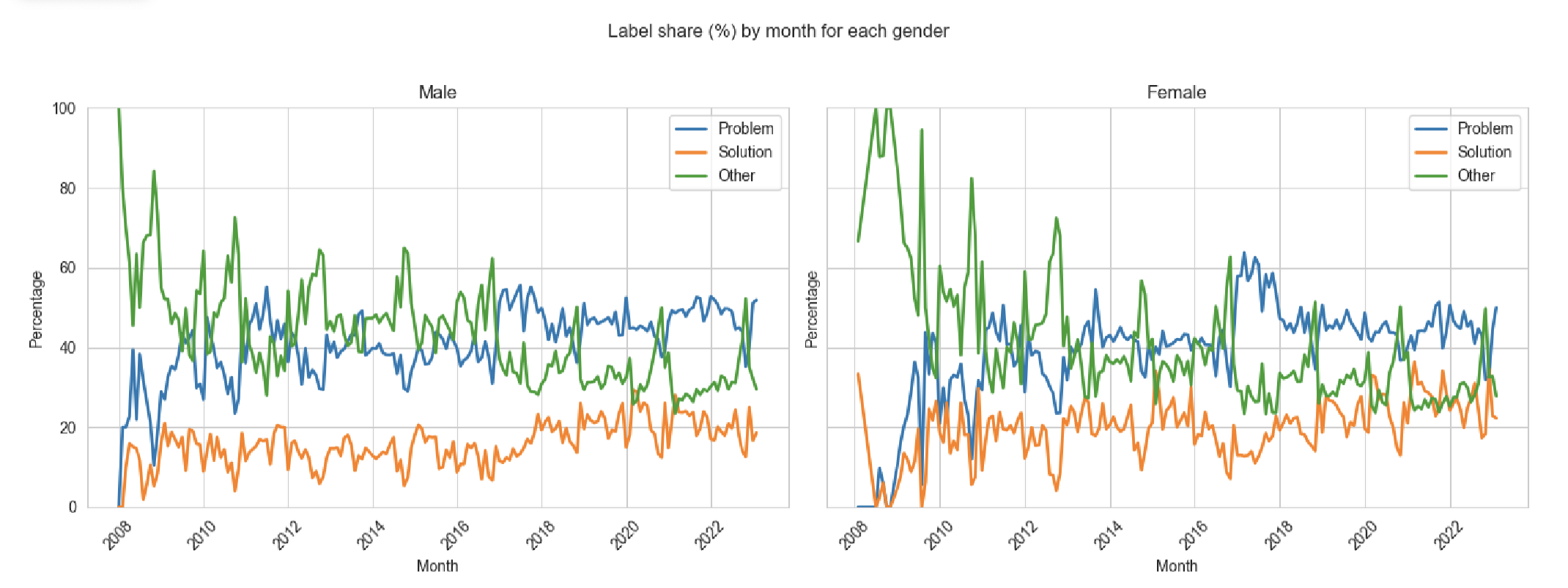}
    \includegraphics[width=\textwidth]{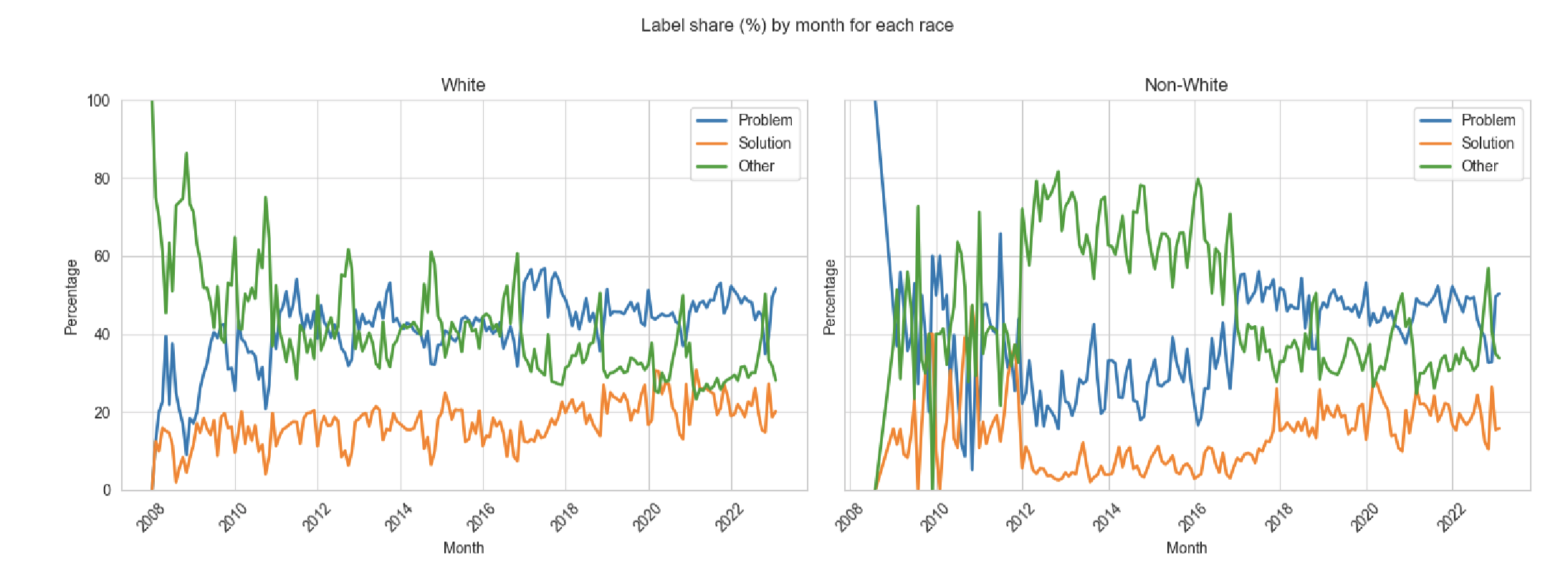}
    \end{center}
  \caption{\label{fig:fig6} Distribution of labels per month by gender and race.}
\end{figure}

Figure~\ref{fig:fig6} show shows the distribution by gender and by race. As above, things change in 2017, with an increase in problem tweets starting in 2017, DJT first term. The only thing that stands out here is non-whites increase in ``other'' 2012-2017 second Obama term. One thing that also stands out  is that the variance seems much higher 2008-2011 or so. This is likely an artifact of the \# of tweets/year as shown in the first plot.

\section{Discussion}
This project investigated computational methods for labeling tweets from US senators as either ``problem'' or ``solution'' oriented, a dichotomy arising from the work of~\citet{Kingdon2003}. Our primary objective is to study the time course of political communications to the public through the Kingdonian lens. We base this study on a large set of social media posts from US senators. At around 1.68 million such posts, expert labeling is challenging. In this study, a team of two experts (AR and MR authors) labeled a subset of 4203 tweets and our objective was to label the remaining tweets with sufficient confidence for downstream analysis. Sufficient in this case is considered to be the level of inter-rater agreement between our two experts. This paper described our computational solution to this problem.

We investigated three supervised methods, and three unsupervised or semi-supervised methods. We found a transformer-based classifier (BERT) was able to classify tweets into problem/solution with over .80 F1 scores, which was thought to be sufficient for downstream tasks. Some examples are shown in Table~\ref{tab:examples}. Other methods included a pseudo-labeling method which gave a modest improvement using a boosting approach, and two zero-shot methods based on large-language models. We found the zero-shot methods, aa logistic regression and XGboost baselines, did not match the BERT-based classifier's results.

\begin{table}
\begin{tabular}{|p{3.2in}|l|l|l|l|}\hline
tweet & \multicolumn{2}{c|}{human label}  & \multicolumn{2}{c|}{BERT-base} \\ \cline{2-5}
& AR & MB & label & confidence \\ \hline
{\em We need @Senate\_GOPs to join us to help small businesses grow \& create jobs} & problem & problem & problem & 0.96 \\ \hline
{\em New @HHSGov report found nearly 1 in 3 people on Medicare was prescribed opioid painkillers last year} & problem & problem & problem & 0.96 \\ \hline
{\em VIDEO: Maria visits GE Aviation in \#Yakima to discuss growing WA's aerospace industry:} & problem &  other&  problem & 0.87\\ \hline
{\em \#Israel faces unique threats but offers amazing opportunity.} & problem & other & problem & 0.96 \\ \hline
{\em This \#MemorialDay, we honor those who gave their lives for \#USA \& pledge to help all vets access health care, education \ job opportunities.} & other & problem & problem & 0.81 \\ \hline
{\em \#NoBonusesForTaxCheats would withhold bonuses from \#IRS employees delinquent on taxes or have documented misconduct.} & solution & solution & solution & 0.92 \\ \hline
{\em \#SecretLaw shouldn’t be used to infringe on ordinary Americans’ civil liberties. @MarkUdall \& I working to end this}  & solution & solution & solution & 0.91 \\ \hline
{\em My \#SIPs legislation is a key first step toward solving real social problems with programs that are proven to work.} & solution & solution & solution & 0.96 \\ \hline
{\em New polls - Gov. Romney up 20 points with veterans. They see the President's lack of leadership. \#LeadingFromBehind} & other & other & other & 0.63 \\ \hline
\end{tabular}
\caption{\label{tab:examples} Examples from TEST (links removed).}
\end{table}

\section{Conclusion}
In this paper, we investigated methods for labeling a large dataset of US senator social media posts (on Twitter). Based on a small set of labeled tweets, we were able to automatically label the remaining 1.68 million tweets with sufficient confidence. We showed some descriptive results of this automatic labeling, with some encouragingly clear transitions over time or ``phase-changes'' that seem correlated with US political elections.

\appendix
\section{Examples}
\section{Further details on pseudo-labeling}
\begin{itemize}
\item Class-Specific Threshold Effects
  \begin{itemize}
    \item Class 3 (Political) Recognition
      \begin{itemize}
        \item At k = 10: Class 3 F1 = 0.67, high recall (0.82), low precision (0.58)
        \item At k = 22.5: Class 3 F1 = 0.74, balanced recall (0.71) and precision (0.77)
        \item At k = 40: Class 3 F1 = 0.65, low recall (0.52), high precision (0.88)
      \end{itemize}
      \item Class 1 \& 2 Differentiation
        \begin{itemize}
          \item Lower k values improved differentiation between Classes 1 and 2 by assigning ambiguous tweets to Class 3.
          \item At k = 22.5, Class 1 F1 improved from 0.71 to 0.78 compared to direct classification
        \end{itemize}
  \end{itemize}
\item Confidence Distribution Patterns, Analysis of confidence distributions revealed interesting patterns:
  \begin{itemize}
    \item Class Separation
      \begin{itemize}
      \item Clear tweets showed strong confidence separation
      \item Ambiguous tweets showed more balanced confidence
        \end{itemize}
    \item Implicit Class 3 Confidence
      \begin{itemize}
      \item Strong Class 3 tweets typically had low conf1 and conf2 scores
      \item The implicit confidence approach effectively captured tweets that didn't fit clearly into Classes 1 or 2
      \end{itemize}
    \item Confidence Gaps
      \begin{itemize}
      \item Gap between highest and second-highest confidence correlated with classification accuracy
      \item Large gaps ($>50\%$) yielded 95\% classification accuracy
      \item Small gaps ($<10\%$) yielded only 68\% classification accuracy
      \end{itemize}
  \end{itemize}
\end{itemize}

\bibliographystyle{apalike}
\bibliography{../../refs}
\end{document}